\documentclass[a4paper,11pt]{article}
\usepackage{pos}

\newlength{\bibitemsep}
\newlength{\bibparskip}
\let\oldthebibliography\thebibliography
\renewcommand\thebibliography[1]{%
  \oldthebibliography{#1}%
  \setlength{\parskip}{\bibitemsep}%
  \setlength{\itemsep}{\bibparskip}%
}

\title{\textit{Fermi}-LAT realtime follow-ups of high-energy neutrino alerts}
 \ShortTitle{\textit{Fermi}-LAT follow-ups of neutrinos}

\author*[a]{S. Garrappa}
\author[b]{S. Buson}
\author[a,c]{A. Franckowiak}
\author[d]{M. Giroletti}
\author[e]{I. Liodakis}
\author{on behalf of the \textit{Fermi}-LAT Collaboration}
\author[d,f]{C. Nanci}

\affiliation[a]{DESY,Platanenallee 6, 15387 Zeuthen, Germany.}
\affiliation[b]{Institut f\"ur Theoretische Physik and Astrophysik, Universit\"at W\"urzburg, D-97074 W\"urzburg, Germany.}
\affiliation[c]{Fakult\"at für Physik \& Astronomie, Ruhr-Universit\"at Bochum, D-44780 Bochum, Germany}
\affiliation[d]{INAF Istituto di Radioastronomia, via Gobetti 101, I-40129 Bologna (Italy)}
\affiliation[e]{Finnish Center for Astronomy with ESO, Quantum, Vesilinnantie 5, University of Turku, FI-20014, Finland}
\affiliation[f]{Department of Physics and Astronomy, University of Bologna, Via Gobetti 93/2 – 40129 Bologna (Italy)}

\emailAdd{simone.garrappa@desy.de}

\abstract{The detection of the flaring gamma-ray blazar TXS 0506+056 in spatial and temporal coincidence with the high-energy neutrino IC-170922A represents a milestone for multi-messenger astronomy. The prompt multi-wavelength coverage from several ground- and space-based facilities of this special event was enabled thanks to  the key role of the \textit{Fermi}-Large Area Telescope (LAT), continuously monitoring the gamma-ray sky. Exceptional variable and transient events, such as bright gamma-ray flares of blazars, are regularly reported to the whole astronomical community to enable prompt multi-wavelength observations of the astrophysical sources. As soon as realtime IceCube high-energy neutrino event alerts are received, the relevant positions are searched, at multiple timescales, for gamma-ray activity from known sources and newly detected emitters positionally consistent with the neutrino localization.
In this contribution, we present an overview of follow-up activities and strategies for the realtime neutrino alerts with the \textit{Fermi}-LAT, focusing on some interesting coincidences observed with gamma-ray sources. We will also discuss future plans and improvements in the strategies for the identification of gamma-ray counterparts of single high-energy neutrinos.}

\FullConference{37$^{\rm{th}}$ International Cosmic Ray Conference (ICRC 2021)\\
		July 12th -- 23rd, 2021\\
		Online -- Berlin, Germany}

\begin{document}
\maketitle

\section{Introduction}
The new era of multimessenger astronomy with neutrinos has significantly boosted the amount of interesting results on possible connections between astrophysical sources and high-energy neutrino observations, with the introduction in 2016 by the IceCube South Pole Neutrino Observatory of a realtime stream of alerts shared with the global astronomical community \cite{Aartsen:2016lmt}. After the detection of a diffuse neutrino flux in 2013 \cite{icecube2013evidence}, the most compelling question was on the origin of this astrophysical component, that remains still unknown.\\
Major challenges in addressing this question come from the limited angular resolution of neutrino detectors and the uneven coverage of the sky by the majority of observing facilities, that work uniquely in pointed observation modes. In addition, the interpretations of observed data with models that include photohadronic ($p\gamma$) and hadronuclear ($pp$) interactions are still highly debated, so there are no clear observational signatures that would identify a hadronic source. Gamma rays are considered one of the most promising signatures for the identification of cosmic-ray accelerators, being produced in the decay of neutral pions that accompany the charged pionic component responsible for the production of neutrinos.\\
The Large Area Telescope (LAT) on board of the \textit{Fermi} Satellite has a key role in the identification of neutrino sources, being sensitive to gamma rays with energies from 20 MeV to greater than 300 GeV \cite{Atwood:2009ez} and with its almost 13 years of operations in survey mode, scanning the entire sky every $\sim$3 hours. \\
The \textit{Fermi}-LAT observations have been crucial for the identification of the so far most interesting coincidence between a high-energy neutrino and an astrophysical source. On September 22 2017, a well-reconstructed neutrino event, with a high probability of being astrophysical and with an estimated energy of $\sim 290$ TeV, was detected in spatial and temporal coincidence with the gamma-ray flaring blazar TXS 0506+056 \cite{IceCube:2018dnn}. Prompt follow-up observations by the \textit{Fermi}-LAT Collaboration \cite{Tanaka:2017atel} triggered a rich multi-wavelength campaign that involved 18 observational facilities unveiling the first compelling evidence of a blazar as neutrino source.
Although it is already known from stacking analyses that the gamma-ray blazar population cannot be solely responsible for the whole diffuse neutrino flux \cite{Aartsen:2016lir}, several blazar candidates have been identified as likely neutrino sources. Among these, the powerful flat spectrum radio quasar (FSRQ) PKS 1502+106 in spatial coincidence with the event IceCube-190730A \cite{garrapa2019ATel_neutrinoPKS1502} and the high-synchrotron-peaked object NVSS J065844+063711 with the alert IceCube-201114A \cite{garrapa2020ATel_neutrino} have been so far the most interesting coincidences observed since the start of the IceCube Realtime Alert Stream 2.0 \cite{Blaufuss:2019fgv} in 2019. Other blazar candidates such as PKS 1424-41 \cite{2016NatPh..12..807K} and GB6 J1040+0617 \cite{2019ApJ...880..103G} have been identified as candidate neutrino sources because of gamma-ray flaring activities in temporal and spatial coincidence with high-energy neutrino events observed  before the start of the IceCube Realtime Program. \\
In this contribution, we present the follow-up strategy of the \textit{Fermi}-LAT team to realtime neutrino alerts, and a summary of all the results that we have published so far by means of rapid publication like Circulars for the Gamma-ray Coordinates Network (GCN) and Astronomer's Telegrams (ATel). In Section \ref{sec:fup} we present the \textit{Fermi}-LAT follow-up strategy for realtime neutrino alerts, with a collection of coincidences observed with known LAT catalog sources. In Section \ref{subsec:pks} and \ref{subsec:nvss} we highlight the most interesting coincidence obtained with LAT observations and in Section \ref{sec:summary} we present the current status and future perspectives.

\section{Follow-up observations with \textit{Fermi}-LAT}
\label{sec:fup}
\subsection{Analysis strategy}
Since the start of the IceCube Realtime Alert Stream in 2016, the \textit{Fermi}-LAT follow-up strategy for neutrino alerts consists of a systematic analysis of the sky region around the neutrino arrival direction, looking for possible transient emission from known sources or new gamma-ray emitters at timescales that range from a single day up to the full set of historical observations obtained by the LAT over the entire mission. \\
We investigate 3 different timescales during a standard follow-up analysis with LAT data (with T$_{0}$ being the neutrino arrival time):
\begin{itemize}
    \item \textbf{1 day before T$_{0}$}: sensitive to the detection of fast, bright transients, down to few-hours duration.
    \item \textbf{1 month before T$_{0}$}: sensitive to the detection of recent transients from the sources of interest.
    \item \textbf{Full-mission data up to T$_{0}$}: Study long-term behavior of LAT catalog sources and detect weak gamma-ray emitters not significantly detected in the LAT catalogs.
\end{itemize}

The choice of the aforementioned timescales for the standard investigation of neutrino alert regions is motivated by a trade-off between the instrument sensitivity and the expected time lag between gamma-ray and neutrino emission from time-dependent studies of blazars (see \cite{2019NatAs...3...88G} for an example applied to the gamma-ray flare of TXS 0506+056 in 2017).\\
After the issue from the Astrophysical Multimessenger Observatory Network (AMON)/GCN stream of the first Notice with the preliminary localization of the neutrino event, the timescales analyses over 1-day and 1-month are performed to check promptly for recent short transients. Since the downlink (and availability to download) of LAT data has a typical delay of up to 3-4 hours, at the time of the issue of the refined localization from IceCube via GCN Circulars after a few hours, the analysis is repeated for the new position at all standard timescales with data up to the neutrino arrival time T$_{0}$. When significant detections are found at short timescales, a lightcurve analysis up to one year prior to T$_{0}$ is performed to characterize the temporal evolution of the source in the short and medium term. In the case of a non-detection at the best-fit position of the neutrino event, upper limits at the 95$\%$ C.L. corresponding to the detection of a source with a power-law model with index 2.0  are reported for each of the time scales.\\
For typical configurations of the LAT follow-up analysis, we select photons from the event class
developed for point-source analyses in the energy range from 100 MeV up to 1 TeV binned into 10 logarithmically spaced energy intervals per decade. We select a region of interest (ROI) of at least 15$\times$15 degrees centered on the neutrino best-fit position, binned in 0.1 deg size pixels. The binning is applied in celestial coordinates using a Hammer–Aitoff projection. We use standard data-quality cuts to select events observed when the detector was in a normal operation mode and remove time periods with the Sun within 15 deg of the ROI center. We perform a maximum-likelihood analysis using the latest available version of the \textit{Fermi}-LAT ScienceTools package (along with the latest IRFs and diffuse models) from the \textit{Fermi} Science Support Center (FSSC) and the Fermipy package \cite{2017ICRC...35..824W}.
\begin{figure*}[h!]
    \includegraphics[width=0.7\linewidth]{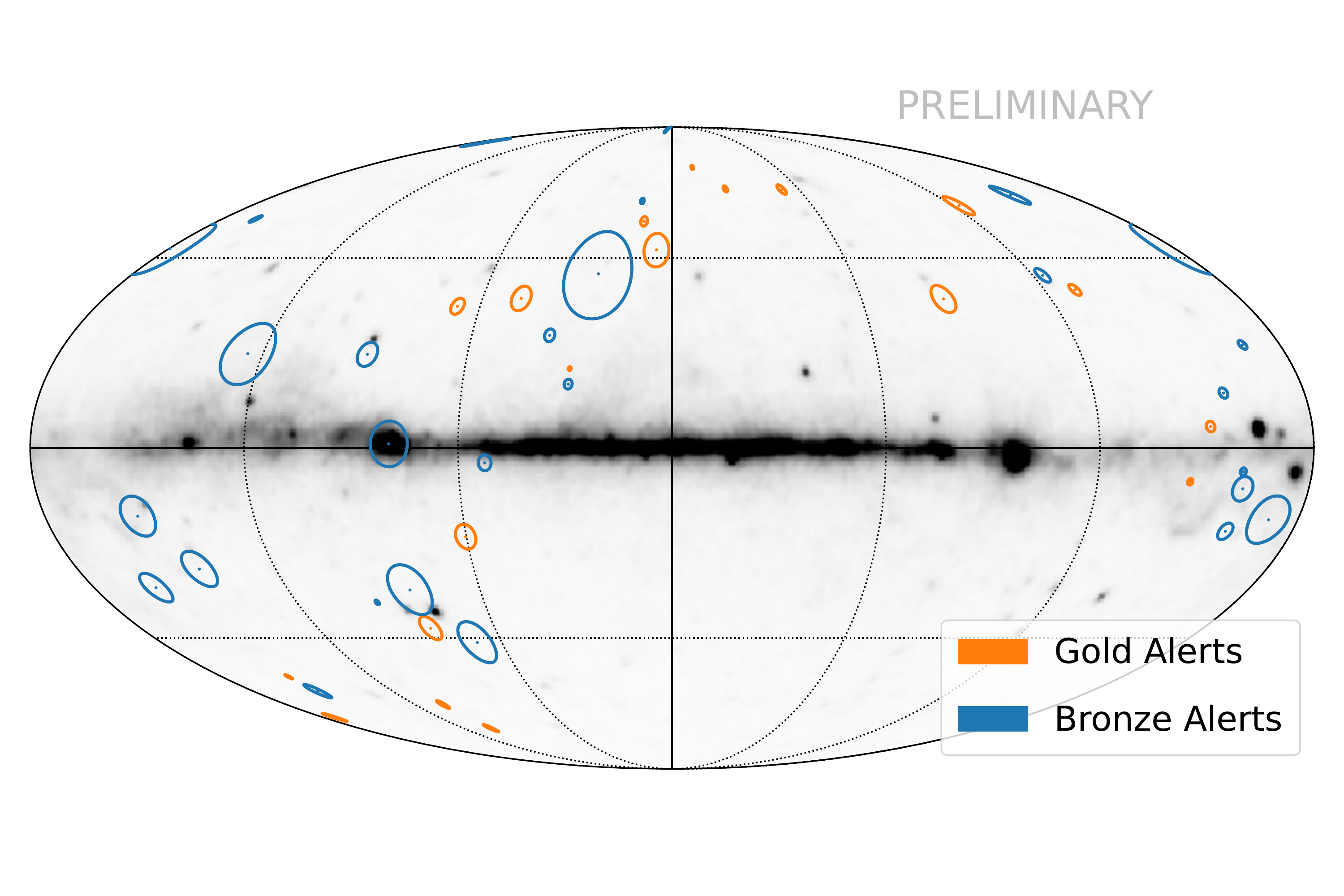}
    \centering
    
    \caption{All-sky map showing the best-fit positions and 90$\%$ containment regions (approximated to circles) of the new IceCube Realtime Alert stream in Galactic coordinates. Gold Alerts are shown in \textit{orange} and Bronze alerts in \textit{blue} (see text for definition of such alerts).}
    \label{fig:skymap}
\end{figure*}

\subsection{Follow-up results}
\label{sec:fup_res}
\begin{figure*}[h!]
    \includegraphics[width=0.5\linewidth]{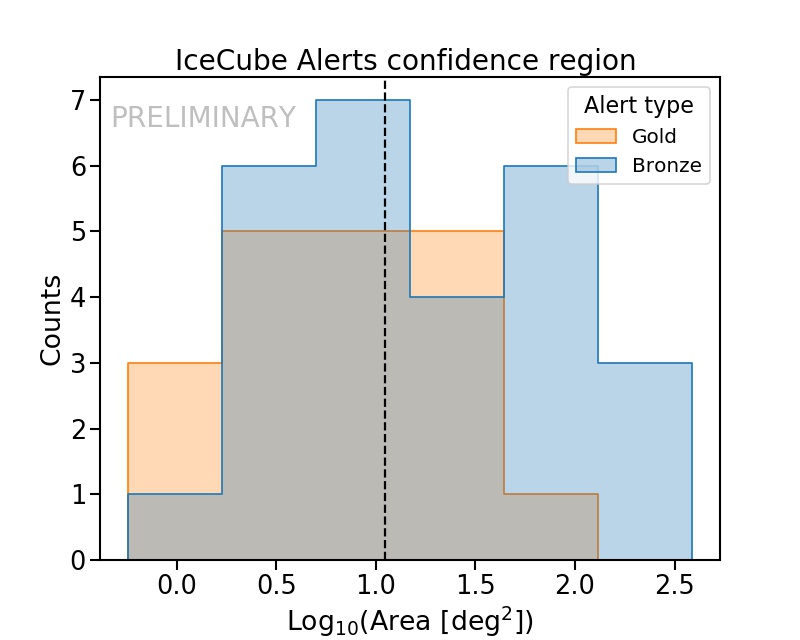}\includegraphics[width=.5\linewidth]{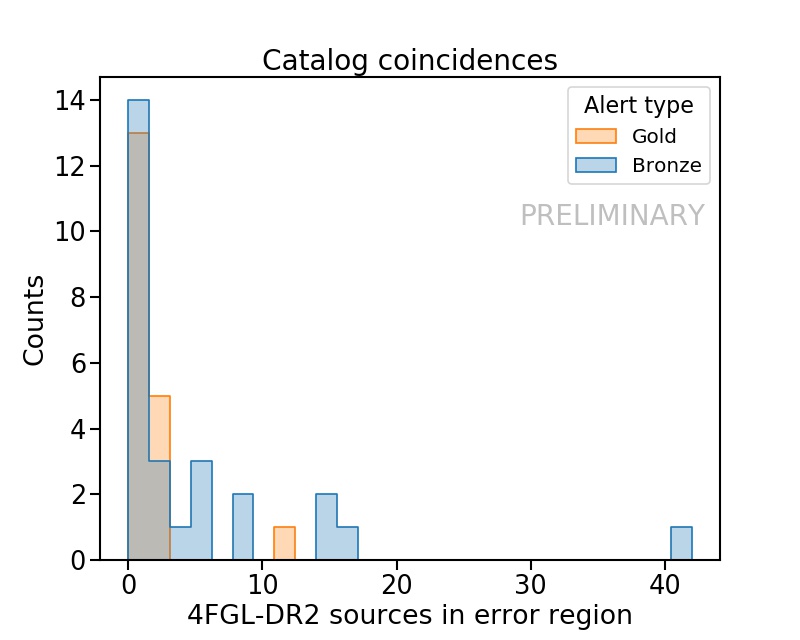}

    \caption{Distribution of 90$\%$ error region extensions  \textit{(left)} and number of 4FGL sources coincident \textit{(right)} divided by alert type. The dashed line in the \textit{(left)} plot shows the median value for the full sample. }
    \label{fig:errors_4fgl}
\end{figure*}

At the time of the writing of this contribution (June 18, 2021) the \textit{Fermi}-LAT  observed all 46 alerts issued by the IceCube Realtime Stream 2.0. Among these, 19 were classified as 'Gold' and 27 as 'Bronze' based on their 'signalness' (average probability of being of astrophysical origin greater than 50$\%$ and 30$\%$, respectively) \ref{fig:skymap}. The left panel of Figure \ref{fig:errors_4fgl} shows the distributions of the neutrino 90$\%$ containment regions for each alert classification, ranging from 0.57 deg$^{2}$ up to 385 deg$^{2}$. The median extension of the full sample of neutrino 90$\%$ containment regions is 11.4 deg$^{2}$ (black dashed line) while it is 5.6 deg$^{2}$ for the 'Gold' sample and 13.4 deg$^{2}$ for the 'Bronze'.\\
We show in the right panel of Figure \ref{fig:errors_4fgl} the number of 4FGL-DR2 \cite{2020arXiv200511208B} sources coincident with the neutrino 90$\%$ containment regions. For 20 events, there are no 4FGL-DR2 sources coincident (43$\%$ of the whole sample), while there is a single 4FGL-DR2 candidate for 7 events. The remaining 19 have several source candidates.\\

In order to select coincidences from relatively well-reconstructed neutrino alerts, we consider the median angular uncertainty of 11.4 deg$^{2}$ observed in the whole alert sample as a selection cut for the study of potentially interesting coincidences. With an average density of AGN objects in the Fourth \textit{Fermi}-LAT Catalog of Active Galactic Nuclei (4LAC \cite{2020ApJ...892..105A}) of $\sim$0.07  deg$^{-2}$ ($\sim$0.12  deg$^{-2}$ for the whole 4FGL-DR2), we still expect a non-negligible rate of random chance coincidences of catalog sources with neutrino alerts, but this selection cut should help to reduce random coincidences from poorly reconstructed events.\\
After the application of this selection cut, 23 alerts are left in the sample (12 'Gold', 11 'Bronze') and only 7 of them have at least one 4FGL-DR2 coincidence, one of them has 2. We list these coincidences in Table \ref{tab:coincidence_table} where we show the neutrino signalness, coincident sources and their 4FGL-DR2 energy flux.\\

\begin{figure*}
    \centering
    \includegraphics[width=0.7\linewidth]{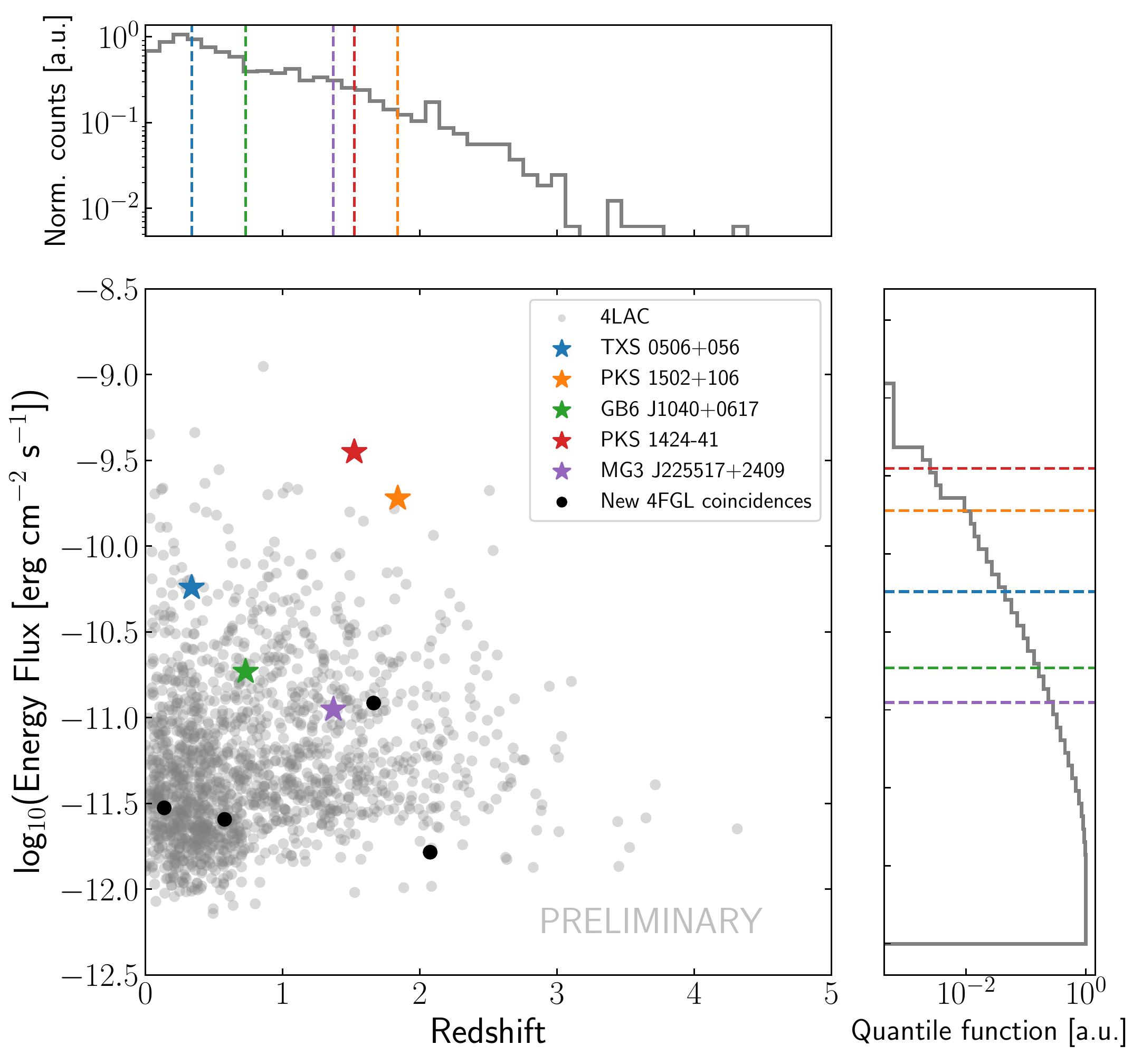}

    \caption{Comparison of candidate neutrino blazars with all blazars in the 4LAC AGN sample adapted from \cite{2020ApJ...893..162F}. Star-shaped markers indicate candidate neutrino blazars already studied in the literature and black circles show the new coincident sources (with identified counterpart) from Table \ref{tab:coincidence_table}.}
    \label{fig:EfluxRedshift}
\end{figure*}

In Figure \ref{fig:EfluxRedshift} we show an updated version of the plot from \cite{2020ApJ...893..162F} with a comparison between the time-integrated gamma-ray energy flux in the energy range from 100 MeV to 100 GeV as a function of redshift for all blazars in 4LAC and the sub-sample of sources coincident with single neutrino events (star markers are for candidate neutrino sources already identified in the literature and black circles are for the additional sub-sample of spatial coincidences from Table \ref{tab:coincidence_table}).
It can be seen in Table \ref{tab:coincidence_table} that some of these neutrino alerts have a low (<50$\%$) probability of being of astrophysical origin, so we should expect that at least half of this sample of coincidences is of random nature due to the atmospheric origin of the detected events. \\
In \cite{2020ApJ...893..162F}, the authors find weak evidence that single high-energy neutrino
emission is correlated linearly with the gamma-ray brightness of the blazars. If these results are confirmed in the future with the increase of the statistics of alerts and coincidences, gamma-ray observations will be crucial in the selection of neutrino source candidates among blazars.

\begin{table}[h!]
\centering
\caption{Selection of 4FGL-DR2 sources coincident with well-reconstructed realtime events from Section \ref{sec:fup_res} }
\begin{tabular}{lllllll}
\hline
\textbf{4FGL Name} & \textbf{Class}\footnotemark[1]         & \textbf{E.Flux [erg cm$^{-2}$ s$^{-1}$]}\footnotemark[2] & \textbf{Redshift}     & \textbf{Event} & \textbf{Type} & \textbf{Sig.} \\ \hline
J1504.4+1029  & FSRQ                      & (1.9 $\pm$ 0.02)$\times$10$^{-10}$      & 1.84                  & IC190730A      & Gold                 & 0.67                \\
J0946.2+0104  & BL Lac                    & (2.55 $\pm$ 0.55)$\times$10$^{-12}$      & 0.577                 & IC190819A      & Bronze               & 0.29                \\
J1003.4+0205  & BCU                       & (1.64 $\pm$ 0.39)$\times$10$^{-12}$      & 2.075                 & IC190819A      & Bronze               & 0.29                \\
J0658.6+0636  & BCU                       & (3.7 $\pm$ 0.73)$\times$10$^{-12}$      &\multicolumn{1}{c}{-}  & IC201114A      & Gold                 & 0.56                \\
J0206.4-1151  & FSRQ                      & (1.22 $\pm$ 0.06)$\times$10$^{-11}$     & 1.663                 & IC201130A      & Gold                 & 0.15                \\
J1342.7+0505  & BL Lac                    & (2.98 $\pm$ 0.49)$\times$10$^{-12}$      & 0.13663               & IC210210A      & Gold                 & 0.65                \\
J1747.6+0324  & \multicolumn{1}{c}{unid.} & (7.03 $\pm$ 0.92)$\times$10$^{-12}$      & \multicolumn{1}{c}{-} & IC210510A      & Bronze               & 0.28     

\end{tabular}

\label{tab:coincidence_table}
\end{table}

\footnotetext[1]{Classification in 4FGL-DR2}
\footnotetext[2]{4FGL-DR2 Energy Flux from 100MeV to 100 GeV}

\subsection{Outstanding coincidences: PKS 1502+106 and IceCube-190730A}
\label{subsec:pks}
On July 30th 2019, IceCube reported the detection of a well reconstructed 'Gold' neutrino event with a $67\%$ probability of being of astrophysical origin~\cite{2019ATel12967....1T}. The only LAT source located within the uncertainty region of IceCube-190730A was the FSRQ PKS 1502+106 (4FGL J1504.4+1029), one of the highest redshift \textit{Fermi}-blazars (z=1.84)~\cite{2010MNRAS.405.2302H}, and the 15th brightest out of 2863 sources in the 4LAC catalog in terms of gamma-ray energy flux at $>100$\,MeV.\\
Despite several flaring activity episodes having been observed from the source during the whole LAT mission, PKS 1502+106 was detected in one of its lowest historical states, with no significant detection up to the 1-week timescale before the neutrino arrival time. The source was observed in a quiescent state for the entire year before the neutrino detection, after having been in a bright, continuous flaring state for almost 4.5 years \cite{garrapa2019ATel_neutrinoPKS1502}. A rich multi-wavelength campaign that was performed for this target, did not show any remarkable activity for the source, except for radio observations from OVRO \cite{2020ApJ...893..162F} where the source was registered at one of its highest flux levels after a slow, long-term increase that started in 2014.\\
Multi-wavelength observations of the source were studied in detail in \cite{2020ApJ...893..162F}  and they were modeled under the hypothesis of hadronic processes in \cite{2021ApJ...912...54R}, obtaining predictions consistent with the observation of a single high-energy neutrino during the gamma-ray quiescient state. Interestingly, \cite{2021MNRAS.503.3145B} and \cite{2020ApJ...894..101P} find that its radio properties also make PKS 1502+106 a valid candidate neutrino source.

\subsection{Outstanding coincidences: NVSS J065844+063711 and IceCube-201114A} 
\label{subsec:nvss}
The 'Gold' alert IceCube-201114A was observed on November 11 2020 and reported with a signalness of 56$\%$ with a well-reconstructed direction. The only LAT source consistent with the 4.56 deg$^{2}$ extension of the 90$\%$ containment region is 4FGL J0658.6+0636 (also present in the third catalog of hard \textit{Fermi}-LAT sources, 3FHL \cite{2017ApJS..232...18A}), associated with the high-synchrotron peaked blazar NVSS J065844+063711. In the follow-up analysis after the issue of the realtime alert, the source was not significantly detected in the LAT data at 1-day and 1-month timescales before the neutrino arrival time \cite{garrapa2020ATel_neutrino}. More extensive investigations of the source revealed the detection in LAT data of VHE emission from the source up to 155 GeV \cite{2020ATel14200....1B} during 12 years of observations. A rich multi-wavelength campaign from radio to IACT observations was coordinated in order to get quasi-simultaneous data of the source right after the neutrino detection. A detailed description of these observations are presented in \cite{deMenezesPoSICRC}.

\section{Summary and perspectives}
\label{sec:summary}
\textit{Fermi}-LAT follow-up observations of realtime high-energy neutrino detections have already identified several interesting candidate counterparts under the hypothesis of a direct connection between gamma-ray and neutrinos in the multi-messenger domain. The \textit{Fermi}-LAT is continuously improving its follow-up strategies towards a faster and more detailed reporting of observations, in order to trigger promptly other multi-wavelength facilities on outstanding target candidates. The team is also involved in active proposals for follow-up radio observations in the European VLBI Network (EVN), optical polarization measurements at the Nordic Optical Telescope (NOT) and RoboPol at Skinakas Observatory,  X-ray observations with \textit{Swift}-XRT and several other collaborations with observing facilities.

\section*{Acknowledgements}
The \textit{Fermi}-LAT Collaboration acknowledges support for LAT development, operation and data analysis from NASA and DOE (United States), CEA/Irfu and IN2P3/CNRS (France), ASI and INFN (Italy), MEXT, KEK, and JAXA (Japan), and the K.A.~Wallenberg Foundation, the Swedish Research Council and the National Space Board (Sweden). Science analysis support in the operations phase from INAF (Italy) and CNES (France) is also gratefully acknowledged. This work performed in part under DOE Contract DE-AC02-76SF00515.
SG and AF acknowledge the support by the Initiative and Networking Fund of the Helmholtz Association. SB acknowledges financial support by the European Research Council for the ERC Starting grant MessMapp, under contract no. 949555.


\begin{thebibliography}{99}
\bibitem{Aartsen:2016lmt}
M.~G.~Aartsen \textit{et al.} [IceCube],
``The IceCube Realtime Alert System,''
Astropart. Phys. \textbf{92} (2017), 30-41
doi:10.1016/j.astropartphys.2017.05.002
[arXiv:1612.06028 [astro-ph.HE]].

\bibitem{icecube2013evidence}
Aartsen, M. G. et al.
\textit{Evidence for high-energy extraterrestrial neutrinos at the IceCube detector
}. Science, 342, 6161, 2013.

\bibitem{Atwood:2009ez}
W.~B.~Atwood \textit{et al.} [\textit{Fermi}-LAT],
``The Large Area Telescope on the \textit{Fermi} Gamma-ray Space Telescope Mission,''
Astrophys. J. \textbf{697} (2009), 1071-1102
doi:10.1088/0004-637X/697/2/1071
[arXiv:0902.1089 [astro-ph.IM]].


\bibitem{IceCube:2018dnn}
M.~G.~Aartsen \textit{et al.},
``Multimessenger observations of a flaring blazar coincident with high-energy neutrino IceCube-170922A,''
Science \textbf{361} (2018) no.6398, eaat1378
doi:10.1126/science.aat1378
[arXiv:1807.08816 [astro-ph.HE]].


\bibitem{Tanaka:2017atel}
Tanaka T. et al. \textit{Fermi-LAT detection of increased gamma-ray activity of TXS 0506+056, located inside the IceCube-170922A error region}. The Astronomer's Telegram, 10791, 1, 2017.

\bibitem{Aartsen:2016lir}
M.~G.~Aartsen \textit{et al.} [IceCube],
Astrophys. J. \textbf{835} (2017) no.1, 45
doi:10.3847/1538-4357/835/1/45
[arXiv:1611.03874 [astro-ph.HE]].

\bibitem[Garrappa et al.(2019)]{garrapa2019ATel_neutrinoPKS1502} Garrappa, S., Buson, S., \& Gasparrini, D.\ 2019, The Astronomer's Telegram, 12972


\bibitem{garrapa2020ATel_neutrino}
Garrappa, S. \& Buson, S. \textit{Fermi-LAT Gamma-ray Observations of IceCube-201114A}. The Astronomer's Telegram, 14188, 1, 2020.

\bibitem{Blaufuss:2019fgv}
E.~Blaufuss \textit{et al.} [IceCube],
PoS \textbf{ICRC2019} (2020), 1021
doi:10.22323/1.358.1021
[arXiv:1908.04884 [astro-ph.HE]].

\bibitem[Kadler et al.(2016)]{2016NatPh..12..807K} Kadler, M., Krau{\ss}, F., Mannheim, K., et al.\ 2016, Nature Physics, 12, 807.

\bibitem[Garrappa et al.(2019)]{2019ApJ...880..103G} Garrappa, S., Buson, S., Franckowiak, A., et al.\ 2019, ApJ, 880, 103.

\bibitem[Gao et al.(2019)]{2019NatAs...3...88G} Gao, S., Fedynitch, A., Winter, W., et al.\ 2019, Nature Astronomy, 3, 88. doi:10.1038/s41550-018-0610-1

\bibitem[Wood et al.(2017)]{2017ICRC...35..824W} Wood, M., Caputo, R., Charles, E., et al.\ 2017, 35th International Cosmic Ray Conference (ICRC2017), 301, 824

\bibitem[Ballet et al.(2020)]{2020arXiv200511208B} Ballet, J., Burnett, T.~H., Digel, S.~W., et al.\ 2020, arXiv:2005.11208

\bibitem[Ajello et al.(2020)]{2020ApJ...892..105A} Ajello, M., Angioni, R., Axelsson, M., et al.\ 2020, ApJ, 892, 105. doi:10.3847/1538-4357/ab791e

\bibitem[Franckowiak et al.(2020)]{2020ApJ...893..162F} Franckowiak, A., Garrappa, S., Paliya, V., et al.\ 2020, ApJ, 893, 162. doi:10.3847/1538-4357/ab8307

\bibitem[Taboada \& Stein(2019)]{2019ATel12967....1T} Taboada, I. \& Stein, R.\ 2019, The Astronomer's Telegram, 12967


\bibitem[Hewett \& Wild(2010)]{2010MNRAS.405.2302H} Hewett, P.~C. \& Wild, V.\ 2010, MNRAS, 405, 2302. doi:10.1111/j.1365-2966.2010.16648.x

\bibitem[Rodrigues et al.(2021)]{2021ApJ...912...54R} Rodrigues, X., Garrappa, S., Gao, S., et al.\ 2021, ApJ, 912, 54. doi:10.3847/1538-4357/abe87b

\bibitem[Britzen et al.(2021)]{2021MNRAS.503.3145B} Britzen, S., Zaja{\v{c}}ek, M., Popovi{\'c}, L. {\v{C}}., et al.\ 2021, MNRAS, 503, 3145. doi:10.1093/mnras/stab589

\bibitem[Plavin et al.(2020)]{2020ApJ...894..101P} Plavin, A., Kovalev, Y.~Y., Kovalev, Y.~A., et al.\ 2020, ApJ, 894, 101. doi:10.3847/1538-4357/ab86bd

\bibitem[Ajello et al.(2017)]{2017ApJS..232...18A} Ajello, M., Atwood, W.~B., Baldini, L., et al.\ 2017, ApJS, 232, 18. doi:10.3847/1538-4365/aa8221

\bibitem[Buson et al.(2020)]{2020ATel14200....1B} Buson, S., Garrappa, S., \& Cheung, C.~C.\ 2020, The Astronomer's Telegram, 14200

\bibitem{deMenezesPoSICRC}
de~Menezes, R., et al. PoS(ICRC2021)955 
\end{thebibliography}
\end{document}